\documentclass[11pt]{article}
\usepackage{latexsym}  % for Symbol fonts
\usepackage{amssymb}
\usepackage{graphicx}
\usepackage{amsmath}
\begin{document}
\hspace*{11cm} {OU-HET-659/2010}

\begin{center}
{\Large\bf Quark Mass Matrix Model for Neutrino Mixing}\footnote{
Talk given at the CTP international conference on Neutrino Physics 
in the LHC Era, 15-19 Nov. 2009, Luxor, Egypt. 
To appear in the Proceedings, Int.J.Mod.Phys.A (2010).}

\vspace{5mm}
{\bf Yoshio Koide}

{\it Department of Physics, Osaka University,  
Toyonaka, Osaka 560-0043, Japan} \\
{\it E-mail address: koide@het.phys.sci.osaka-u.ac.jp}
\end{center}

\begin{abstract}
For the purpose of deriving the 
observed nearly tribimaximal neutrino mixing, a possible 
quark mass matrix model is investigated based on a 
yukawaon model. 
The neutrino mass matrix is related to the up-quark mass
matrix.       
Five observable quantities (2 up-quark mass ratios and 3 
neutrino mixing parameters) are excellently fitted by 
two parameters( one parameter in the up-quark mass matrix
and another one in the neutrino mass matrix).   
Also, the CKM mixing parameters and down-quark mass ratios 
are given under the other 2 parameters. 
\end{abstract}

%%%%%%%%%%%%%%%%%%%%%%%%%%%%%%%%%%%%%%%%%%%%%%%%%%%%
\section{What is a yukawaon model?}

Usually, the tribimaximal neutrino mixing\cite{tribi} 
has been derived from scenarios based on discrete symmetries. 
In contrast to the conventional approach, 
in the present work, we will try another approach without
such discrete symmetries.
For the purpose of deriving the 
observed nearly tribimaximal neutrino mixing, a quark 
mass matrix model is investigated based on the so-called 
``yukawaon" model.   
Five observable quantities (2 up-quark mass ratios and 3 
neutrino mixing parameters) are excellently fitted by 
two parameters.   Also, the Cabibbo-Kobayayashi-Maskawa (CKM) 
mixing parameters and down-quark mass ratios are given 
under the other 2 parameters. 

First, let us give a short review of the yukawaon model.
In the standard model of quarks and leptons, 
the Yukawa coupling constants $Y_f$ are fundamental constants 
in the theory.  
Even if we assume flavor symmetries in order to
reduce the number of the fundamental constants $(Y_f)_{ij}$,
some of $(Y_f)_{ij}$ will still remain as fundamental 
constants in the theory. 
If we make a multi-Higgs extension in which 
Higgs scalars have flavor quantum numbers, we will encounter
some of troubles, e.g. a flavor changing neutral current 
problem, unwelcome behavior of the SU(2)$_L$ $\beta$-function, 
and so on.
In the yukawaon model, effective Yukawa coupling constants
$Y_f^{eff}$ are given by vacuum expectation values (VEVs) of 
new gauge singlet scalars $Y_f$:
$$
Y_f^{eff} =\frac{y_f}{\Lambda} \langle Y_f\rangle ,
\eqno(1)
$$
where $\Lambda$ is a scale of an effective theory which is 
valid at $\mu \leq \Lambda$, and we assume 
$\langle Y_f\rangle \sim \Lambda$.
We refer to the fields $Y_f$ as 
``yukawaons"\cite{yukawaon} hereafter. 
Those VEVs can, in principle, be calculated dynamically, 
although, at present, the dynamics for the 
yukawaons is not yet established.

However, we have a hint for this dynamics: 
Recently, Sumino\cite{Sumino09} 
has proposed a very interesting model for the charged 
lepton mass relations by assuming U(3)$\times$O(3)
flavor gauge symmetries. 
In his model, the electro-magnetic corrections for 
the charged lepton  masses are exactly cancelled by 
the family gauge  interactions, so that the charged 
lepton mass  relation for the running masses are given 
by the  same form as that for the pole masses.

In the present work, differently from Sumino's scenario, 
we will adopt a conventional scenario given in a series 
of our works: 
(i) Masses which we deal with are not ``pole masses",  but 
``running masses".
(ii) We assume a global O(3) flavor symmetry, which is completely 
broken at $\mu \sim \Lambda$. 
(ii) We consider $\Lambda \sim 10^{14-16}$ GeV, and 
$Y_f^{eff}$ evolve as those in the standard model below the scale 
$\Lambda$. 
(iv) In order to obtain VEV relations, we will use SUSY vacuum 
conditions, so that SUSY are still unbroken for $\mu \sim \Lambda$.

In the present work, we assume an O(3) flavor symmetry.
Would-be Yukawa interactions are given by
$$
H_{Y}= \sum_{i,j} \frac{y_u}{\Lambda} u^c_i(Y_u)_{ij} {q}_{j} H_u  
+\sum_{i,j}\frac{y_d}{\Lambda} d^c_i(Y_d)_{ij} {q}_{j} H_d 
$$
$$
+\sum_{i,j} \frac{y_\nu}{\Lambda} \ell_i(Y_\nu)_{ij} \nu^c_{j} H_u  
+\sum_{i,j}\frac{y_e}{\Lambda} \ell_i(Y_e)_{ij} e^c_j H_d +h.c. 
+ \sum_{i,j}y_R \nu^c_i (Y_R)_{ij} \nu^c_j .
\eqno(2)
$$ 
All of the yukawaons $Y_f$ belong to 
${\bf 3}\times{\bf 3}=({\bf 5}+{\bf 1})_S$ of O(3). 
In order to distinguish each $Y_f$ from others, 
we assume a U(1)$_X$ symmetry (i.e. ``sector charge")
in addition to the O(3) symmetry, and we have assigned 
U(1)$_X$ charges as $Q_X(Y_f)=x_f$, $Q_X(f^c)=-x_f$ and
$Q_X(\nu^c)=2 x_\nu$.
(The SU(2)$_L$ doublet fields $q$, $\ell$, $H_u$ and $H_d$
are assigned to sector charges $Q_X=0$.) 
For the neutrino sector, we assume 
$Q_X(\nu^c) =Q_X(e^c)$, so that the yukawaon $Y_e$ can also 
couple to the neutrino sector as $(\ell Y_e \nu^c) H_u$ 
instead of $(\ell Y_\nu \nu^c) H_u$ in Eq.(2).
We do not need a yukawaon $Y_\nu$ in the present model.

Then, we obtain VEV relations as follows: 
(i) We give an O(3) and U(1)$_X$ invariant superpotential for 
yukawaons $Y_f$. 
(ii) We solve SUSY vacuum conditions $\partial W/\partial Y_f=0$.
(iii) Then, we obtain VEV relations among $Y_f$. 
For example, we have assume the following superpotential
$$
W_e = \lambda_e {\rm Tr}[\Phi_e \Phi_e \Theta_e] 
+\mu_e {\rm Tr}[Y_e \Theta_e] + W_\Phi, 
\eqno(3)
$$
where we have assumed $Q_X(\Phi_e) = \frac{1}{2}Q_X(Y_e) =
-\frac{1}{2}Q_X(\Theta_e)$ and the term $W_\Phi$ has been
introduced in order to determine a VEV spectrum 
$\langle \Phi_e \rangle$ completely. 
Then, from the  SUSY vacuum condition
$$
\frac{\partial W}{\partial \Theta_e} = \lambda_e \Phi_e \Phi_e +
\mu_e Y_e =0 ,
\eqno(4) 
$$
we obtain a VEV relation
$$
\langle Y_e \rangle =
-\frac{\lambda_e}{\mu_e} \langle \Phi_e\rangle  \langle \Phi_e\rangle . 
\eqno(5)
$$ 
The scalar $\Theta_e$ does not have a VEV, i.e. 
$\langle \Theta_e \rangle =0$.
Therefore, terms which include more than two of $\Theta_e$
do not play any physical role, so that we do not consider 
such terms in the present effective theory. 

%%%%%%%%%%%%%%%%%%%%%%%%%%%%%%%%%%%%%%%%%%%%%%%%%%%%%%%%%%%
\section{Why quark mass matrix model for neutrino mixing?}     

In a previous work\cite{Koide-O3-PLB08} on the neutrino mixing 
based on a yukawaon model, for the seesaw-type neutrino mass matrix,  
$M_\nu \propto \langle Y_\nu\rangle \langle Y_R\rangle^{-1} 
\langle Y_\nu\rangle^T$, the author has obtained the following VEV 
relations  
$$
\langle Y_R\rangle \propto  \langle Y_e\rangle 
\langle \Phi_u\rangle + \langle\Phi_u\rangle \langle Y_e\rangle 
\eqno(6)
$$ 
together with $\langle Y_u\rangle \propto \langle\Phi_u\rangle 
\langle\Phi_u\rangle$. 
Here, the relation (6) has been derived from a superpotential
$$
W_R = \mu_R {\rm Tr}[Y_R \Theta_R]
 +{\lambda_R}
 {\rm Tr}[(Y_e \Phi_u+\Phi_u Y_e) \Theta_R] .
 \eqno(7)
$$
As a result, the neutrino mass matrix is given by a form
$$
\langle M_\nu \rangle_e \propto \langle Y_e\rangle_e 
\left\{ \langle Y_e\rangle_e \langle \Phi_u\rangle_e 
+ \langle\Phi_u\rangle_e \langle Y_e\rangle_e 
\right\}^{-1} \langle Y_e\rangle_e ,
\eqno(8)
$$
where 
$$
\langle \Phi_u\rangle_u \propto {\rm diag}
(\sqrt{m_u}, \sqrt{m_c} , \sqrt{m_t}) .
\eqno(9)
$$
Here, $\langle A \rangle_f$ denotes a form of a VEV matrix
$\langle A \rangle$ in the diagonal basis of $\langle Y_f\rangle$
(we refer to it as $f$ basis).
We can obtain a form $\langle \Phi_u\rangle_d = V(\delta)^T 
\langle \Phi_u\rangle_u V(\delta)$ from the definition 
of the CKM matrix $V(\delta)$, but we do not know an explicit 
form of $\langle \Phi_u\rangle_e$.
Therefore, in a previous work\cite{Koide-O3-PLB08}, we put an ansatz 
$$ 
\langle \Phi_u\rangle_e = V(\pi)^T  \langle \Phi_u\rangle_u
V(\pi)
\eqno(10)
$$ 
by supposing $\langle \Phi_u\rangle_e \simeq 
\langle \Phi_u\rangle_d$, and we obtained excellent predictions
of the neutrino oscillation parameters 
$\sin^2 2\theta_{atm}=0.995$, $|U_{13}|=0.001$ and 
$\tan^2\theta_{solar}=0.553$, 
without assuming any discrete symmetry.

However, there is no theoretical ground for the ansatz (10)
for the form  $\langle \Phi_u\rangle_e$.
The purpose of the present work is to investigate a 
quark mass matrix model in order to predict neutrino
mixing parameters on the basis of a yukawaon model (2), 
without such the ad hoc ansatz,  because if we give a quark 
mass matrix model where mass matrices $(M_u, M_d)$ are given 
on the $e$ basis, then,  we can obtain the form 
$\langle \Phi_u\rangle_e$ by using a transformation 
$$
\langle \Phi_u\rangle_e = U_u \langle \Phi_u\rangle_u U_u^T ,
\eqno(11)
$$
where $U_u$ is defined by $U_u^T M_u U_u= M_u^{diag}$. 

%%%%%%%%%%%%%%%%%%%%%%%%%%%%%%%%%%%%%%%%%%%%%%%%%%%%%%
\section{Yukawaons in the quark sector}

We assume a superpotential\cite{Mnu_PLB09} in the quark sector:
$$
W_q = \mu_u {\rm Tr}[Y_u \Theta_u] +
\lambda_u {\rm Tr}[\Phi_u \Phi_u \Theta_u] 
+ \mu_u^X {\rm Tr}[\Phi_u\Theta_u^X] 
+\mu_d^X {\rm Tr}[Y_d \Theta_d^X]
$$
$$
+ \sum_{q=u,d} \frac{\xi_q}{\Lambda} {\rm Tr}[\Phi_e 
(\Phi_{X} + a_q E ) \Phi_e \Theta_q^X] .
\eqno(12)
$$
From SUSY vacuum conditions $\partial W/\partial \Theta_u =0$, 
$\partial W/\partial \Theta_u^X =0$ and 
$\partial W/\partial \Theta_d^X =0$, we obtain 
$\langle Y_u\rangle \propto \langle\Phi_u\rangle 
\langle\Phi_u\rangle$, 
$$
M_u^{1/2}\propto 
\langle \Phi_u \rangle_e \propto \langle \Phi_e \rangle_e 
\left(
\langle \Phi_X \rangle_e + a_u \langle E \rangle_e
\right) \langle \Phi_e \rangle _e ,
\eqno(13)
$$ 
$$
M_d \propto 
\langle Y_d \rangle_e \propto \langle \Phi_e \rangle_e 
\left(
\langle \Phi_X \rangle_e + a_d \langle E \rangle_e
\right) \langle \Phi_e \rangle _e ,
\eqno(14)
$$
respectively. 
Here, $\langle \Phi_X \rangle_e$ and 
$\langle E \rangle_e$ are given by
$$
\langle \Phi_X \rangle_e \propto 
X \equiv \frac{1}{3} \left(
\begin{array}{ccc}
1 & 1 & 1 \\
1 & 1 & 1 \\
1 & 1 & 1 
\end{array} \right), \ \ \ \ \ 
\langle E \rangle_e \propto 
{\bf 1} \equiv  \left(
\begin{array}{ccc}
1 & 0 & 0 \\
0 & 1 & 0 \\
0 & 0 & 1 
\end{array} \right) .
\eqno(15)
$$
(Note that the VEV form $\langle \Phi_X \rangle_e$ 
breaks the O(3) flavor symmetry into S$_3$.)
Therefore, we obtain quark mass matrices
$$
M_u^{1/2} \propto  M_e^{1/2} \left( {X} + 
a_u   {\bf 1} \right) M_e^{1/2} , \ \ \ 
M_d \propto  M_e^{1/2} \left( {X} + 
a_d e^{i\alpha_d}  {\bf 1} \right) M_e^{1/2} ,
\eqno(16)
$$ 
on the $e$ basis.
Note that we have assumed that the O(3) relations are valid
only on the $e$ and $u$ bases, so that $\langle Y_e\rangle$ 
and $\langle Y_u\rangle$ must be real. 

A case $a_u \simeq -0.56$ can give a reasonable up-quark
mass ratios 
$$
\sqrt{\frac{m_{u1}}{m_{u2}}}=0.043, \ \ \  
\sqrt{\frac{m_{u2}}{m_{u3}}}=0.057 ,
\eqno(17)
$$
which are in favor of
the observed values\cite{q-mass}
$\sqrt{{m_{u}}/{m_{c}}}=0.045^{+0.013}_{-0.010}$, and
$\sqrt{{m_{c}}/{m_{t}}}=0.060\pm 0.005$ at $\mu=m_Z$. 

\section{Yukawaons in the neutrino sector}

However, the up-quark mass matrix (16) failed to give reasonable 
neutrino oscillation parameter values although it can 
give reasonable up-quark mass ratios.  
Therefore, we will slightly modify the model (6) 
in the neutrino sector.

Note that the sign of the eigenvalues of $M_u^{1/2}$
given by Eq.(16) is $(+, -,+)$ for the case $a_u \simeq -0.56$. 
If we assume that the eigenvalues of $\langle \Phi_u\rangle_u$ 
must be positive, so that $\langle \Phi_u\rangle_u$ in Eq.(6) 
is replaced as $\langle \Phi_u\rangle_u \rightarrow 
\langle \Phi_u\rangle_u \cdot {\rm diag}(+1,-1,+1)$,
then, we can obtain successful results except for 
$\tan^2 \theta_{solar}$, i.e. predictions 
$\sin^2 2\theta_{atm}=0.984$ and $|U_{13}|=0.0128$ and
an unfavorable prediction $\tan^2 \theta_{solar}=0.7033$
(see predicted values with $\xi=0$ in Table 1).

When we introduce a new field $P_u$ with a VEV
$$
\langle P_u \rangle_u \propto {\rm diag}(+1,-1,+1) ,
\eqno(18)
$$
we must consider an existence of 
$P_u Y_e \Phi_u+\Phi_u Y_e P_u$ in addition to 
$Y_e P_u \Phi_u+\Phi_u P_u Y_e$, 
because they have the same U(1)$_X$ charges.
Therefore, we modify Eq.(7) into
$$
W_R = \mu_R {\rm Tr}[Y_R \Theta_R]
 + \frac{\lambda_R}{\Lambda}
\left\{ {\rm Tr}[(Y_e P_u \Phi_u+\Phi_u P_u Y_e) \Theta_R] 
\right.
$$
$$
\left.
+\xi {\rm Tr}[(P_u Y_e \Phi_u +\Phi_u Y_e P_u) \Theta_R] \right\},
\eqno(19)
$$
which leads to VEV relation 
$$
Y_R \propto Y_e P_u \Phi_u + \Phi_u P_u Y_e 
+\xi (P_u Y_e \Phi_u + \Phi_u Y_e P_u) .
\eqno(20)
$$

%\section{Numerical Results }

We list numerical results from the model (20) in Table 1.
The results at $a_u \simeq -0.56$ are excellently in favor
of the observed neutrino oscillation parameters
$\sin^2\theta_{atm}=1.00_{-0.13}$ \cite{MINOS} and
$\tan^2\theta_{solar}=0.469^{+0.047}_{-0.041}$ \cite{SNO08} 
by taking a small value of $|\xi|$.

%\begin{table}[ph]
\begin{center}
\begin{footnotesize}
%\tbl{
{ Table 1.  Predicted values for the neutrino 
oscillation parameters} 

\vspace{1mm}

\begin{tabular}{cccc} \hline
$\xi$ & $\sin^2 2\theta_{atm}$ & 
$\tan^2 \theta_{solar}$ & $|U_{13}|$ \\ \hline
0  & 0.9848 & 0.7033 & 0.0128 \\ \hline
$+0.004$  &  0.9825 & 0.4891 & 0.0123 \\
$+0.005$ &  0.9819 & 0.4486 & 0.0122 \\ 
$+0.006$ &  0.9812 & 0.4123 & 0.0120 \\ \hline
$-0.0011$ & 0.9897 & 0.4854 & 0.0142 \\
$-0.0012$ & 0.9900 & 0.4408 & 0.0143 \\
$-0.0013$ & 0.9904 & 0.4008 & 0.0144 \\
\hline
\end{tabular} 
\end{footnotesize}
\end{center}                       
%\end{table}

Also, we can calculate the down-quark sector.
We have two parameters $(a_d, \alpha_d)$ in the 
down-quark sector given in Eq.(16). 
As seen in Table 2, 
the results are roughly reasonable, although 
$|V_{ub}|$ and $|V_{td}|$ are somewhat larger than 
the observed  values.   
Those discrepancies will be improved in
future version of the model.

%\begin{table}[ph]
\begin{center}
\begin{footnotesize}
%\tbl{
{Table 2.  Predicted values for the CKM 
mixing parameters} 

\vspace{1mm}
\begin{tabular}{ccccccccc} \hline
$a_u$ & $a_d$ & $\alpha_d$ & $|m_{d1}/m_{d2}|$ & 
$|m_{d2}/m_{d3}|$ & $|V_{us}|$ & $|V_{cb}|$ & 
 $|V_{ub}|$ & $|V_{td}|$ 
\\ \hline
$-0.56$ & $-0.620$  &$4^\circ$ & 0.1078 & 0.0273 & 
0.2035 & 0.0666 & 0.0101 & 0.0178 \\
$-0.56$ & $-0.625$  &$6^\circ$ & 0.0783 & 0.0313 & 
0.2187 & 0.0818 & 0.0123 & 0.0190 \\
$-0.56$ & $-0.630$  &$8^\circ$ & 0.0542 & 0.0362 & 
0.2222 & 0.0977 & 0.0146 & 0.0194 \\
$-0.58$ & $-0.630$  &$2^\circ$ & 0.1959 & 0.0195 & 
0.2272 & 0.0448 & 0.0088 & 0.0163 \\ \hline
\end{tabular}
\end{footnotesize}
\end{center} 
                      
%\end{table}

%%%%%%%%%%%%%%%%%%%%%%%%%%%%%%%%%%%%%%%%%%%%%%%
\section{Concluding remarks}

The numerical results are summarized in Table 3.
The model predicts masses and mixings for quarks and
leptons with a quit few parameters: 
Five observable quantities (2 up-quark mass ratios and 3 
neutrino mixing parameters) are excellently fitted by 
two parameters.   Also, the CKM mixing parameters and 
down-quark mass ratios are given under the other 2 parameters. 

In the present paper, we have not mentioned how to obtain 
the VEV spectrum $\langle\Phi_e\rangle$.
The VEV value $\langle\Phi_e\rangle$ plays an essential 
role in this model: The value determine the charged lepton 
mass spectrum $\langle Y_e \rangle$ through the relation (5),
the quark mass spectra $(M_u, M_d)$ through the relations
(16), and the neutrino mass matrix $M_\nu$ through the 
relation (20). 
In the present paper, we have used the observed values
$(\sqrt{m_e}, \sqrt{m_\mu}, \sqrt{m_\tau})$ as the
values of $\langle\Phi_e\rangle_e = {\rm diag}
(v_1,v_2,v_3)$.
For a possible mechanism of $\langle\Phi_e\rangle$,
for example, see Ref.\cite{Sumino09,e-spec10}.

\vspace{1mm}

%\begin{table}[ph]
\begin{center}
\begin{footnotesize}
%\tbl{
{Table 3.  Summary table} 

\vspace{1mm}

\begin{tabular}{ccccccc} \hline
Sector & Parameters & Predictions \\ \hline
      &       &  $\sin^2 \theta_{atm}$ \ \ \  $\tan^2 \theta_{solar}$
\ \ \  $|U_{13}|$ \ \\
$M_\nu$ & $\xi=+0.0005$ & $0.982$ \ \ \ \ \ 
$0.449$ \ \ \ \ \ \ \ \ $0.012$ \\ 
  & $\xi=-0.0012$ & $0.990$ \ \ \ \ \ \ 
$0.441$ \ \ \ \ \ \ \ \ $0.014$ \\ 
\cline{2-3}
$M_u^{1/2}$ & $a_u=-0.56$ & $\sqrt{\frac{m_u}{m_c}}=0.0425$ \ \ 
$\sqrt{\frac{m_c}{m_t}}=0.0570$ \\ \cline{2-3}
     & two parameters & 
 5 observables: fitted excellently \\
\hline
$M_d$ & $a_d e^{i\alpha_d}$ &  $\sqrt{\frac{m_d}{m_s}}$, \ 
$\sqrt{\frac{m_s}{m_b}}$, \ $|V_{us}|$, \ $|V_{cb}|$, \ $|V_{ub}|$, \ 
$|V_{td}|$ \\ \cline{2-3}
    & two parameters   & 6 observables: not always excellent \\ 
\hline
\end{tabular}
\end{footnotesize}
\end{center}
%\end{table}

Also, we have not discussed neutrino masses.
In the present model, we can add a term 
$(y'_R/\Lambda) \nu^c Y_e Y_e \nu^c$ to the would-be 
Yukawa interactions (2) under the O(3) and U(1)$_X$ 
symmetries, so that we can always fit the observed neutrino 
mass ratio $R\equiv \Delta m^2_{21}/\Delta m^2_{32}$ 
by adjusting the parameter $y'_R/y_R$ suitably. 
In other words, there is no predictability as far as
the ratio $R$ is concerned. 

Here, we would like to give some comments on the 
present works. 

\noindent 
(i)  We have obtained a nearly tribimaximal  mixing
without assuming any discrete symmetry for the
 neutrino mass matrix, but note that we have assumed
S$_3$ symmetry in the quark sector.

\noindent 
(ii)  Note that it is essential that the quark mass matrices
are given in the $e$ basis in which the charged lepton 
mass matrix takes a diagonal form.
We think that the $e$ basis  has a specific and 
fundamental status in the theory.  
For example, in the $e$ 
basis, the charged lepton masses  satisfy a simple 
relation, and the quark matrices take a simple form.  

\noindent 
(iii) How can we detect a signature of the yukawaon model?
In the present paper, we have assumed that the energy 
scale $\Lambda$ is of the order of $10^{14}$ GeV, so 
that most yukawaon effects will be invisible\cite{invisible}. 
A yukawaon model with a lower energy scale $\Lambda$
is a future task.  

\noindent 
(iv) In the present scenario, an O(3) global symmetry
has been assumed as a flavor symmetry.
However, Sumino\cite{Sumino09} has recently assumed 
a U(3) gauge symmetry in his model. 
His scenario is very attractive. 
 Is family symmetry global or gauged?  
The symmetry is O(3) or U(3)?
At present, those are open questions.  

The present approach will shed new light on 
unified understanding of the masses and mixings.  
At least, the present model will provide a suggestive
hint on a unification model for quarks and leptons.

% ****************************************************************************
% BIBLIOGRAPHY AREA
% ****************************************************************************

\begin{footnotesize}
% IF YOU DO NOT USE BIBTEX, USE THE FOLLOWING SAMPLE SCHEME FOR THE REFERENCES
% ----------------------------------------------------------------------------

\end{footnotesize}

%%%%%%%%%%%%%%%%%%%%%%%%%%%%%%%%%%%%%%%%%%%%%%%%
\end{document}